\documentclass[amsmath,amssymb, aps, bibtex, pre,reprint,shortbibliography
]{revtex4-2}
\usepackage{verbatim}
\usepackage{hyperref} 
\usepackage{graphicx}
\usepackage{dcolumn}
\usepackage{bm}
\usepackage{amsmath}
\usepackage{enumerate}
\usepackage{color}

\newcommand{\beq}{\begin{equation}}
\newcommand{\eeq}{\end{equation}}
\newcommand{\bal}{\begin{aligned}}
\newcommand{\eal}{\end{aligned}}
\newcommand{\be}{\begin{equation}}
\newcommand{\ee}{\end{equation}}
\newcommand{\BE}{\begin{eqnarray}}
\newcommand{\EE}{\end{eqnarray}}

\begin{document}

\title{Navigating Uncertainty: Risk-Averse vs. Risk-Prone Strategies in Populations Facing Demographic and Environmental Stochasticity}

\author{Rub\'en Calvo}
 \affiliation{Instituto Carlos I de F\'isica Te\'orica y Computacional and Departamento de Electromagnetismo y F\'isica de la Materia, Facultad de Ciencias, Universidad de Granada, 18071 Granada, Spain} 
\author{Miguel A. Mu\~noz}%
 \affiliation{Instituto Carlos I de F\'isica Te\'orica y Computacional and Departamento de Electromagnetismo y F\'isica de la Materia, Facultad de Ciencias, Universidad de Granada, 18071 Granada, Spain} 
\author{Tobias Galla}
\affiliation{%
Instituto de F{\' i}sica Interdisciplinar y Sistemas Complejos IFISC (CSIC-UIB), 07122 Palma de Mallorca, Spain
}%

\date{\today}

\begin{abstract}
  Strategies aimed at reducing the negative effects of long-term uncertainty and risk are common in biology, game theory, and finance, even if they entail a cost in terms of mean benefit. Here, we focus on the single mutant's invasion of a finite resident population subject to fluctuating environmental conditions. Thus, the game-theoretical model we analyze integrates environmental and demographic randomness, i.e., the two leading sources of stochasticity and uncertainty. We use simulations and mathematical analysis to study if and when strategies that either increase or reduce payoff variance across environmental states can enhance the mutant fixation probability. Variance aversion implies that the mutant pays insurance in terms of mean payoff to avoid worst-case scenarios. Variance-prone or gambling strategies, on the other hand, entail specialization, allowing the mutant to capitalize on transient favorable conditions, leading to a series of ``boom-and-bust'' cycles. Our analyses elucidate how the rate of change of environmental conditions and the shape of the probability distribution of possible states affect the possible most convenient strategies. We  discuss how our results relate to the bet-hedging theory, which aims to reduce fitness variance rather than payoff variance. We also describe the analogies and differences between these similar yet distinct approaches.
\end{abstract}

\maketitle

\section{\label{sec: Introduction}Introduction}

Biological organisms frequently live in variable and uncertain environments \cite{King2007, Simons2011}. For instance, annual plants may encounter  wet or dry years \cite{SegerBrockmann1987, Philippi1989}, microbial populations may experience temporary nutrient depletion or exposure to antibiotics \cite{Grimbergen2015}, amphibian species may contend with fluctuating temperatures and predator pressure \cite{Crump1981}, and so forth. To overcome these challenges, organisms must evolve strategies to cope with uncertainty.

Responses to uncertainty and variability can broadly be divided into variance-averse (also called ``risk-averse'') and variance-prone (or ``risk-prone'') strategies \cite{Haaland2019}. Variance-averse strategies prioritize stability across time over the possibility of intermittent large gains. 
These actions are designed to minimize the impact of uncertainty and variability, even if it requires sacrificing some potential rewards in the process. On the other hand, variance-prone strategies are characterized by a willingness to accept higher levels of uncertainty and risk in pursuit of potentially higher rewards, even if they come with a higher probability of losses. Variance-averse strategies are often associated with ``generalist'' individuals who perform similarly under diverse circumstances, and variance-prone ones are typical of specialists, meaning that an organism can be highly adapted to some environment, but performs very poorly in others. Whether a particular individual (or species) chooses a variance-averse or variance-prone strategy might depend on factors such as individual risk preferences, the nature of the uncertainty, and the level of accessible information \cite{Camerer2011}.

Within this context, the concept of bet-hedging was first considered almost fifty years ago \cite{Lewontin1969, Gillespie1974, Gillespie1975, Slatkin1974} to refer to a class of strategies that lower the variance of ``fitness" ---i.e. long-term population growth--- even at the expenses of reducing the mean  fitness \cite{Lewontin1969,Childs2010,Kokko,Venable}.  
In this way, bet-hedging strategies aim at maximizing the mean geometric fitness rather than the arithmetic mean \cite{Simons2011,Haaland2020,Hidalgo1}. 
The focus on the geometric mean stems from the intrinsically multiplicative nature of reproduction processes; maximizing the geometric mean is equivalent to minimizing the effects of eventual worst-case scenarios on long-term growth, as briefly discussed in the Appendix \ref{app: Arithmetic vs geometric}. Notably, the celebrated Kelly criterion \cite{kelly1956} in gambling and investment management relies on the same idea \cite{rotando1992,Tal,Lacoste1,Lacoste2}.

Even though there is some debate about the strength of the actual empirical evidence \cite{Simons2011}, variance-averse strategies such as bet-hedging are thought to be employed by several organisms, including viruses, bacteria, insects, and plants. 
\cite{Childs2010,Venable,olofsson_bet-hedging_2009,Grimbergen2015,Hidalgo2}. 
Here, we mainly focus on conservative bet-hedging strategies, where a single phenotype aims to minimize its fitness fluctuations across generations. However, it is important to note that there are also \emph{diversifying} forms of bet-hedging, where a population splits into two or more different phenotypes to reduce population-level fitness variability across generations.

Since the seminal works by Gillespie \cite{Gillespie1974,Gillespie1975}, a significant portion of the literature on strategies for coping with uncertainty has focused on scenarios where environmental or ``extrinsic" fluctuations are the only source of randomness. The population dynamics are otherwise assumed to be deterministic. However, some studies have also 
considered finite populations as a source of ``demographic" stochasticity  \cite{Proulx,Leibler,Hidalgo1}.  In particular, it has been shown that populations can employ variance-averse strategies to cope with the ``intrinsic" uncertainty arising from finite-size fluctuations \cite{Leibler}, for example in small populations colonizing new territory \cite{Villa}.  

\begin{figure*}[htbp]
\centering
    \includegraphics[width=1.0\textwidth]{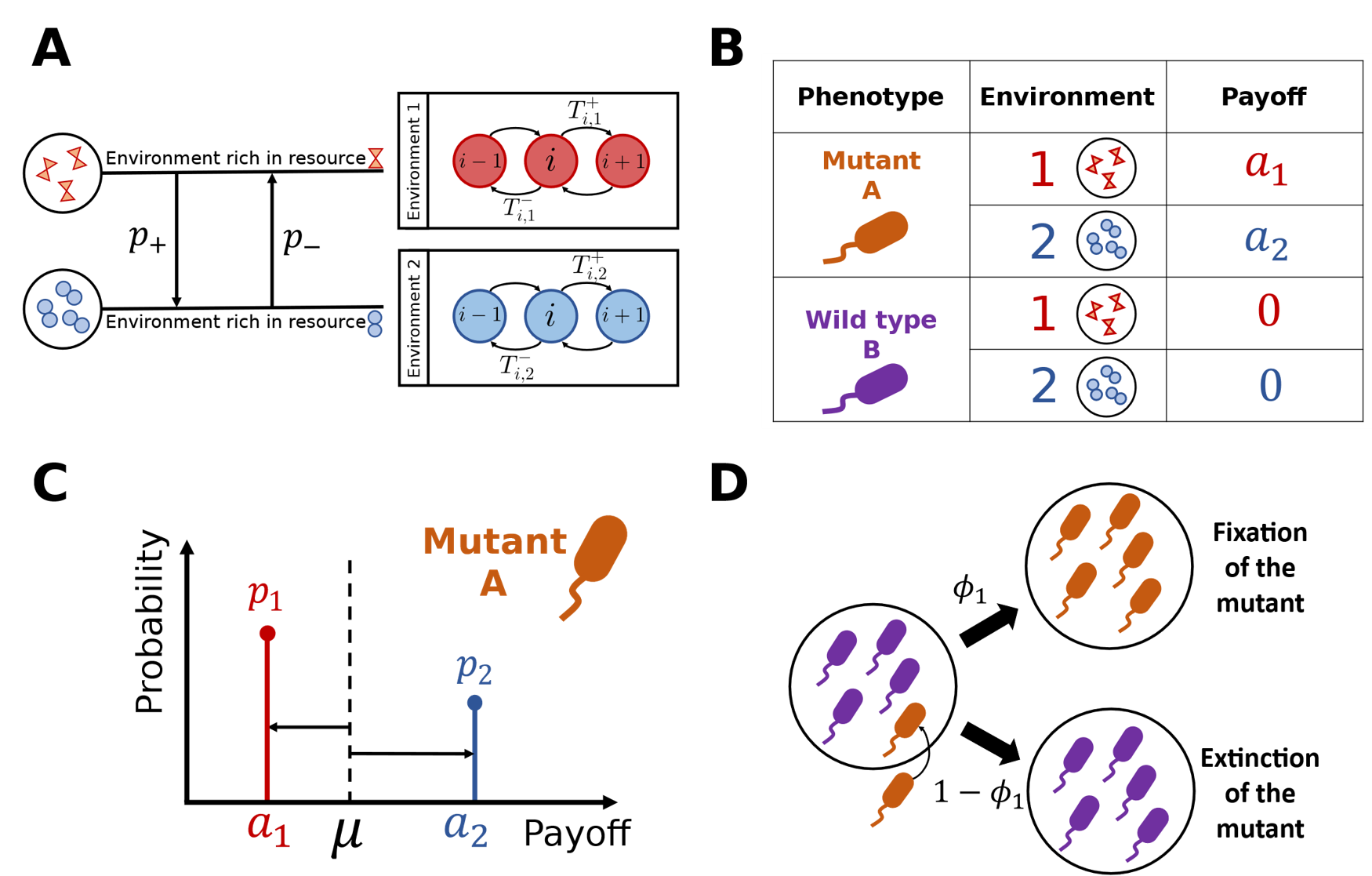}
   \caption{{\bf Illustration of the model setup.} Panel A sketches the two environmental states, $\omega_1$ and $\omega_2$; e.g., each one rich in a particular resource. The switching probabilities (per discrete time unit) from one to the other are given by $p_\pm$, respectively. In either environment, the community evolves through discrete birth-death events with transition probabilities $T_{i,\omega}^\pm$, where $i$ counts the number of mutant individuals and $\omega$ is the current environmental state. Panel B shows the payoff of each particular type (resident or mutant) depending on the current environmental conditions. The wild type (B) has a constant payoff (which can be interpreted as being a generalist, already adapted to the changing environment). Instead, the mutant (A) may have different payoffs, $a_1$ and $a_2$, respectively, in each environment. Panel C shows the probability distribution of payoff values for the mutant, which depends on the distribution of possible environmental conditions. Panel D illustrates the two possible outcomes of evolutionary invasion experiments: either the mutant invade the population and becomes fixated, which happens with a \emph{fixation probability} $\phi_1$,  or the mutant becomes extinct, which happens with complementary probability $1-\phi_1$.}
    \label{fig:model_illustration}
\end{figure*}

Here, we take a broader perspective to explore the conditions under which different strategies become beneficial under environmental and demographic fluctuations. To tackle this, we propose and analyze a simple individual-based model within a game-theoretical framework \cite{nowak_book_2016,Traulsen_2008} for a finite population exposed to randomly changing environmental conditions. In particular, we consider the situation in which a single mutant (or migrant) individual is placed within an established ``resident" or ``wild-type" population. The population dynamics unfold via a succession of stochastic discrete birth-and-death events, whose probabilities are controlled by the ``payoffs" of residents and mutants, respectively \cite{nowak_book_2016,Traulsen_2008}.  The payoffs are considered to be the same across same-type individuals, either resident or mutants, but variable in time since they depend on changing environmental conditions. This implies that any variance-averse strategy would tend to be of the ``conservative" rather than the ``diversifying" type \cite{Kokko,Childs2010}.

 Resident and mutant types can temporarily coexist, however, given the finite population size, the mutant will eventually perish or fixate. The success of the mutant can thus be quantified by its fixation probability \cite{CK,Proulx,Traulsen_2008,ewens2004}. The main question we aim to answer is whether it can be beneficial for the mutant  to shift to a more variance-averse or more variance-prone strategy to increase its fixation probability.

Before proceeding, let us emphasize that ``payoff" ---central to our model and, in general, to game theory--- and ``fitness" ---commonly used in ecological literature including that on bet-hedging--- are related but distinct concepts (see Section \ref{sec: real bet-hedging}). The fitness of an organism is usually identified with reproductive success \cite{crow_kimura}. Payoffs, instead, typically denote the immediate benefit or utility obtained by an individual or strategy from a particular action and can be affected by various factors such as resource availability, competition, cooperation, and the behaviors of other individuals
\cite{Haaland2019}.  Importantly, payoffs do not necessarily translate into an immediate reproductive advantage, i.e. into fitness in a direct way. Actually, in mathematical terms, the mapping from the payoff function to fitness is often nonlinear \cite{Traulsen_Shoresh_Nowak,Haaland2019,Haaland2020}, so that a variance-adverse/prone strategy in terms of one of these concepts could not be so in terms of the other. 
For this reason, here we use the term ``variance-averse" (respectively ``variance-prone'') strategies to describe generically approaches aimed at decreasing (respectively, enhancing) either payoff variance or fitness variance. When there could be ambiguity we use terms such as ``payoff-variance-averse" or ``fitness-variance-averse'.

As we show in what follows, when the mutant's mean payoff is larger than the resident's, it is typically convenient to adopt a more (payoff) variance-averse strategy to avoid risks, even if this is at the expense of reducing its mean payoff. On the other hand, when the mutant's mean payoff is smaller than the resident's, under some circumstances, it may be convenient for the mutant to shift to a more (payoff) variance-prone strategy, allowing it to  specialize in capitalizing on transient advantages offered by particularly favorable environments, even if its mean payoff is further reduced.
However, as we demonstrate, these scenarios  crucially depend on the shape of the probability distribution of possible environmental states and on the time scale of environmental change. For example, slowly changing environments favor the emergence of variance-prone strategies in which the mutant experiences a series of boom-and-bust cycles, which enable it to enhance its fixation probability when favorable conditions come. Finally, shifting the attention from payoffs to fitness, we also make contact with the theory of bet-hedging and the possible emergence of risk-prone strategies in this context.

\section{\label{sec:model}Model definition and setup}

\subsection{Demographic dynamics}
We consider a fixed-size finite population consisting of $N$ individuals, each of which is either a ``mutant" ($A$) or a resident ``wild-type" ($B$).
The constant size assumption is made for analytical convenience, and in line with the standard setting in evolutionary game theory \cite{traulsen_jcc_hauert_2005,Traulsen_2008,nowak_book_2016}
We do not allow for multiple mutant strains to be present simultaneously. Any individual in the population may ``interact'' with any other individual, i.e., the population is well-mixed. Thus, the state of the population is fully described by the number $i$ of type-$A$ individuals (so that the number of type-$B$ individuals is  $N-i$). The population evolves in discrete time steps, $t=0,1,2,\dots$. 
At each step, a single event occurs, involving the asexual reproduction of one individual, giving rise to an offspring of the same type, and the removal (death) of another individual.  The performance of types $A$ and $B$, relative to each other, is governed by their payoffs, $\pi_A$ and $\pi_B$, respectively. For simplicity, we concentrate on frequency-independent selection, where the payoffs can be affected by the environment, but not by the composition of the population \cite{nowak_book_2016,Traulsen_2008}.  

Specifically, at each time step, two individuals are randomly selected from the population: one for reproduction and the other for potential death. If both individuals are of the same type (either  $A$ or $B$) no change occurs.  If, on the other hand, the first individual is of type $A$ (resp. $B$) and the other of type $B$ (resp. $A$)  then, with probability $g^+(\pi_A-\pi_B)$ (resp. $g^-(\pi_A-\pi_B)$) a type-$A$ (resp. type-$B$) individual is added to the community and a type-$B$ (resp. type-$A$) individual is removed from it. 

Among the many possible ways to define these probabilities in terms of payoffs, typically non-linear  \cite{Traulsen_2008,bladon_evolutionary_2010}, we choose the so-called Fermi process:
\be
\label{eq:g_fermi}
g^\pm(\Delta\pi)=\frac{1}{1+\exp\left(\mp \beta \Delta\pi\right)},
\ee
where $\Delta\pi=\pi_A-\pi_B$, where the  parameter $\beta\geq 0$ controls the strength of selection. For large values of $\beta$, the individual with the larger payoff reproduces with very high probability; for $\beta=0$, the dynamics are neutral, i.e., the fitness of all individuals is identical, $g^+=g^-=1/2$, independently of the payoffs. For $0<\beta<\infty$,  $g^+(\pi_A - \pi_B)$ is an increasing function of its argument (while $g^-$ is a decreasing function)  so that, the larger the payoff difference $\pi_A - \pi_B$, the greater the probability that type $A$ reproduces, and the other way around.

At a population level, this defines a one-step-process with transition probabilities $T_i^+$ (from a state with $i$ mutants to one with $i+1$) and $T_i^-$ (from $i$ to $i-1$ mutants) \cite{nowak_evolutionary_2006,Traulsen_2008} given by:
\be 
\label{eq:fermi}
T_i^\pm = \frac{i}{N}\frac{N-i}{N}g^\pm(\pi_A -\pi_B).
\ee
The monomorphic states $i=0$ (extinction of the mutant) and $i=N$ (fixation) are absorbing. If one of these states is reached, no further dynamics can occur. Absorption at one of these states is the inevitable fate of any finite population with the birth-death process in Eq.~(\ref{eq:fermi}), as illustrated in Figure~\ref{fig:model_illustration}~D. We call $\phi_1$ the fixation probability of a single invading mutant; i.e. the probability that absorption is reached at $i=N$ (as opposed to $i=0$) if the dynamics are started from $i=1$.

\subsection{Stochastic environments}
The payoffs $\pi_A$ and $\pi_B$ in Eq.  (\ref{eq:fermi}) are assumed to be affected by an external environment, whose state is denoted $\omega$ and changes in time.  As illustrated in Figure ~\ref{fig:model_illustration}~A, we assume that the environment switches between two discrete states, $\omega=1, 2$. This models situations such as the switching between dry and wet years, feast-famine cycles in resource abundances for bacterial populations, etc. The environment switches between the two states via a Markovian process such that if the environment is in state $\omega=1$ (resp. $\omega=2$) then it changes to state $\omega=2$ (resp. $\omega=1$) with probability $p_+$ (resp. $p_-$) in the next step.  Thus, the mean proportions of time spent in each state are
\be\label{eq:p1p2} p_1=\frac{p_-}{p_++p_-},\;\;\; p_2=\frac{p_+}{p_++p_-}. \ee
The payoffs in state $\omega$ are denoted $\pi_{A,\omega}$ and $\pi_{B,\omega}$, respectively.

\subsection{Variance-averse and variance-prone mutant strategies}

Given that only payoff differences matter for the birth-death dynamics, we set the payoff of the resident (type B) to a constant reference value in both environments, $\pi_{B,\omega=1}=\pi_{B,\omega=2}=0$. In other words, the resident type is assumed to be a generalist, relatively well adapted to all conditions.
Instead, the mutant (type A)  is assumed to have different payoffs for the two environments, that we denote $\pi_{A,\omega=1} =a_1$ and $\pi_{A,\omega=2}= a_2$, respectively (see Figure \ref{fig:model_illustration} B).  

In what follows we label environments such that state $\omega=2$ is always more beneficial to the mutant than environment $\omega=1$. We also assume that the switching probabilities $p_\pm$ are beyond the control of the mutant and resident species. That is to say, the statistics and time scale of the environmental switching are determined externally. For any given environment, as specified by $p_\pm$, we interpret the mutant payoffs $a_1$ and $a_2$ as the consequences of the mutant's strategic choice. In other words, we assume the mutant species has some power to control their payoffs.  Our goal is to compare the probability of reaching fixation of two possible different mutant species, characterized by payoffs $(a_1,a_2)$ and $(a_1',a_2')$. Increasing simultaneously both $a_1$ and $a_2$ will always benefit the mutant. However, assuming that there is some kind of tradeoff, we exclude trivial cases of this type and we ask if and when it can be beneficial for the mutant to increase (or lower) its payoff variance across environments, i.e. become a specialist (resp. more generalist), even if this occurs at the expense of lowering its (arithmetic)  mean payoff.

Before proceeding, let us fix the notation. The mean mutant payoff $\mu$ and the payoff variance $\sigma^2$ across environmental states are given by (see Figure~\ref{fig:model_illustration}~C)
\begin{eqnarray}
\label{eq:mu and sigma} \mu&=&a_1 p_1+a_2 p_2, \nonumber \\
\sigma^2 &=&(a_1-\mu)^2 p_1+(a_2-\mu)^2 p_2.
\end{eqnarray}
Conversely, for future convenience, we can express $a_1$ and $a_2$ in terms of $\mu$ and $\sigma$:
\be
\label{eq:inversion}
a_1=\mu-\sigma\sqrt{\frac{p_2}{p_1}}, \;\;\; ~~~ a_2=\mu+\sigma\sqrt{\frac{p_1}{p_2}}. 
\ee 
Suppose we want to compare two different mutant strains. A change of mutant strategy leads to a change in mean payoff, $\Delta \mu$, and in payoff variance, $\Delta\sigma^2$. To exclude trivial scenarios, we only focus on strategies that do not increase mean payoff, i.e., we require $\Delta\mu\leq 0$. Payoff variance-averse strategies are mutant strategies that reduce the payoff variance ($\Delta\sigma^2<0$). Payoff variance-prone strategies, in contrast, increase the variance ($\Delta\sigma^2>0$). We will also refer to the latter type of strategy as {\em gambling strategies}. We further distinguish between situations in which variance-prone strategies are {\em ``safe"} or {\em ``risky"}, respectively.  We speak of a safe scenario when the more beneficial environment ($\omega=2$) is also the more likely state, i.e., when $p_2>p_1$. Conversely, a strategy is risky if the most beneficial environment ($\omega=2$) is the more unlikely, $p_2<p_2$.

 In our analyses, we quantify the success of any particular mutant's strategy in terms of its fixation probability as described next.

\section{Fixation probability under different environmental conditions: Analytical results}

\subsection{General formalism}

The theory of one-step processes without environmental noise is well established, and the probability of a mutant reaching fixation is well known \cite{kampen_stochastic_2007, ewens2004, Traulsen_2008, nowak_book_2016}. The fixation probability for a single initial mutant $A$ is

\be\label{eq:phi} \phi_1=\frac{1}{\sum_{k=1}^{N-1}\prod_{j=1}^k \gamma_j}, \ee where $\gamma_j \equiv T_j^-/T_j^+$. With complementary probability $1-\phi_1$, the mutant becomes extinct. 
For example, in the case of neutral selection, $T_j^-=T_j^+$,  one has $\phi_1 = 1/N$ \cite{nowak_evolutionary_2006, Traulsen_2008, voter1, voter2}, a value that we will use as a reference.

Calculating the fixation probability when the environment switches between two  states is much more complex, but this problem was tackled in \cite{ashcroft_fixation_2014}. The fixation probability of a single mutant starting from the environmental state $\omega$, $\phi_{1,\omega}$, can be shown to be
\be
\label{eq:S1} 
\phi_{1,\omega}=\frac{1-p_{\bar \omega}}{\Delta} v_{1,\omega}-\frac{p_\omega}{\Delta} v_{1,\bar\omega}, 
\ee 
where $\Delta=1-p_+-p_-$,  $\bar \omega$ denotes the environmental state opposite to $\omega$  (i.e., $\bar 1=2$ and $\bar 2=1$), and where the $2N$ variables $v_{i,\omega}$ are the solution of the linear system 
\be
\label{eq:S2} 
v_{i+1,\omega}=\gamma_{i,\omega} v_{i,\omega}+\frac{1}{T_{i,\omega}^+}\frac{p_\omega}{\Delta}\sum_{j=1}^i (v_{j,\omega}-v_{j,\bar\omega}),
\ee 
with $i=1,\dots,N$, $\gamma_{i,\omega}\equiv T_{i,\omega}^-/T_{i,\omega}^+$,
and are normalized  for both environments
\be
\label{eq:S3}
1=\sum_{i=1}^N v_{i,\omega},\;\;(\omega=1,2). 
\ee 

Finding analytical solutions for these equations is a daunting task, and numerical solutions are only feasible for relatively small population sizes $N$ (in a forthcoming Section we will show numerical results for populations of up to $N=100$ individuals).  However, there are two limiting cases: the fast-switching limit --in which environmental switching occurs on a very short time scale-- and the opposite slow-switching limit --in which the switching events are so rare that the environment effectively remains fixed over many  generations-- for which closed expressions can be found as described in the next two subsections.

\begin{figure*}[htbp]
\centering
    \includegraphics[width=1.0\textwidth]{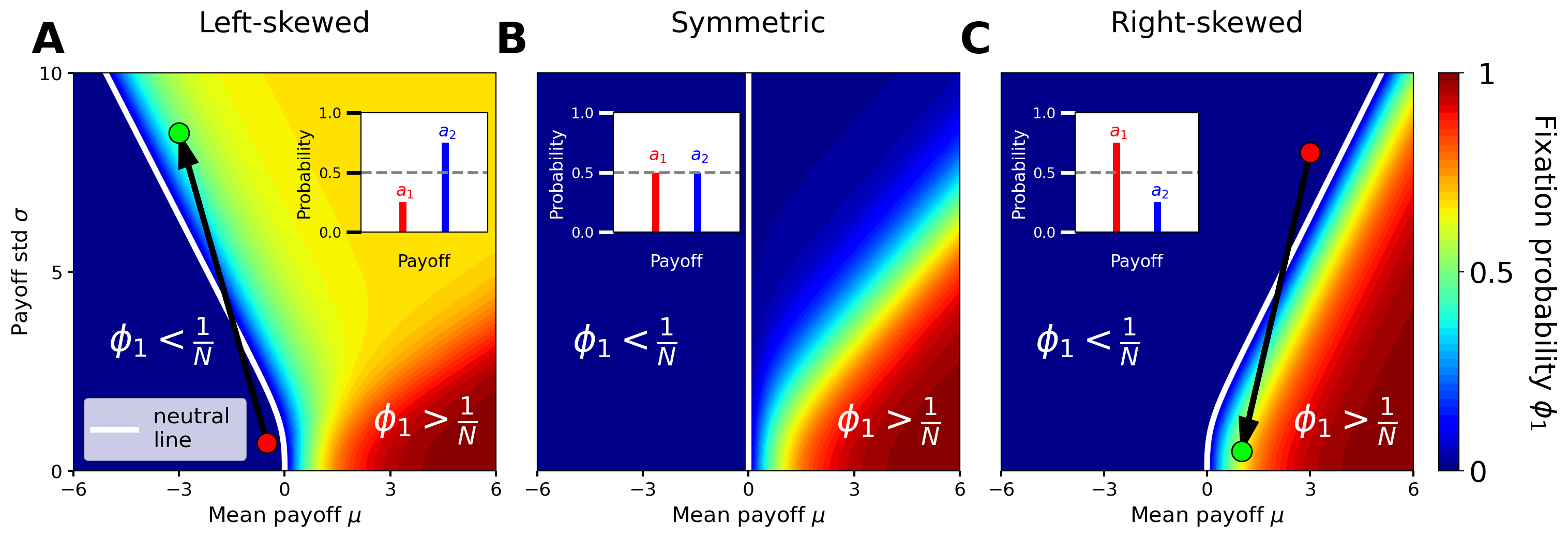}
\caption{{\bf Fixation probability in the fast-switching limit for different payoff distributions.} The line of neutral selection indicates locations in the $(\mu,\sigma)$-plane along which a single invading mutant $A$, has fixation probability $\phi_1=1/N$. Panel A (corresponding to $p_1=0.25$ and  $p_2=0.75$, as illustrated in the inset) shows that the neutral line is tilted to the left. A mutant with a higher payoff variance and lower mean (green dot) can be more likely to invade than a mutant with higher mean payoff and lower variance (red dot). This indicates the possibility of gambling. We note that some other lines of constant $\phi_1$ lean to the right in the region $\mu>0$, thus there is also space for variance-averse strategies. In panel B ($p_1=p_2=0.5$), the neutral line is vertical. The only way to achieve a fixation probability higher than neutral is by increasing the average payoff. In panel C ($p_1=0.75,\: p_2=0.25$), the line of neutral selection tilts to the right, so that a mutant with a lower mean payoff $\mu$ (green dot) is more likely to fixate than a mutant with a higher mean fitness (red dot), provided the first is subject to lower payoff fluctuations than the second. In this scenario, variance-averse strategies are beneficial.}
    \label{fig:neutral_line}
\end{figure*}

\subsection{Fast-switching limit}\label{sec:fast_switching}

The fast-switching limit characterizes a scenario in which environmental switching occurs on a much shorter time scale than the evolution of the population. In our discrete-time model, this means that the switching probabilities $p_\pm$ are sufficiently high so that the number of environmental switches per generation ($N$ birth-death events) is much larger than one (the fastest possible, ultra-fast, dynamics occurs if the environment is randomly chosen at each demographic step). As shown in detail in Appendix \ref{app:fixation_probs}, in the limit of fast switching, the system can be approximated by a birth-death process, with effective transition rates given by averages over the distribution of environmental states:
\be\label{eq: effective transition rates}
T^{\pm}_{i,\text{eff}}=\frac{i}{N}\frac{N-i}{N}
\sum_{\omega} p_\omega  g^\pm(a_\omega).
\ee
The fixation probability for this birth-death process can thus be calculated using Eqs.~(\ref{eq:phi}), where $T_i^\pm$ is to be replaced by $T^{\pm}_{i,\text{eff}}$ and where $\gamma_{i,{\rm eff}}=T^{-}_{i,\text{eff}}/T^{+}_{i,\text{eff}}$. Moreover, given the frequency-independent nature of the model,  $\gamma_{i,\rm eff}$ does not depend on $i$ and is called simply $\gamma$ from now on. Hence, using Eq.~(\ref{eq:phi}) one has
\be\label{eq:phi_1_eff}
\phi_1= \frac{1}{\sum_{k=0}^{N-1}\gamma^k}=\frac{1-\gamma}{1-\gamma^N}.
\ee
In the limit of neutral selection, $\beta\to 0$, the functions $g^\pm$ converge to step functions and $\gamma\to 1$. Using L'H\^opital's rule in Eq.~(\ref{eq:phi_1_eff}), one recovers the  fixation probability of a single mutant under neutral selection, $\lim_{\beta\to 0}\phi_1=1/N$ \cite{voter1,voter2}. Moreover, given that $\phi_1$ is a decreasing function of $\gamma$ for fixed $N$,  $\phi_1=1/N$ if and only if $\gamma=1$. 

Solving the condition $\gamma=1$ in terms of the mean payoff and its variance leads to a \emph{neutral selection line} in ($\mu,\sigma$) space, which ---as shown in  Appendix~\ref{app:neutral_line}---  takes the form
\be\label{eq:neutral_line}
\mu_{\rm neutral}(\sigma)=-\sigma\sqrt{\frac{p_1}{p_2}}-\frac{1}{\beta}\ln{f(\sigma, p_1)},
\ee
where
\BE\label{eq:f}
f(\sigma,p_1)&=&\left(p_1-\frac{1}{2}\right)\left(\frac{1}{\Gamma(\sigma,p_1)}-1\right)\\
&&\hspace{-6em}+\frac{1}{2\Gamma(\sigma,p_1)}\sqrt{4(\Gamma(\sigma,p_1)-1)^2p_1(p_1-1)+(\Gamma(\sigma,p_1)+1)^2},\nonumber
\EE
with
\begin{equation}\label{eq:gamma_main}
\Gamma(\sigma,p_1)=\exp\left(\beta\sigma\left[\sqrt{{p_1/p_2}}+\sqrt{{p_2/p_1}}\right]\right).
\end{equation}
For $p_1=p_2$, i.e. for a symmetric distribution, the expression reduces to $\mu_{\rm neutral}=0$, independently of $\sigma$. This describes the vertical line in Figure~\ref{fig:neutral_line}B. Only mutants with a mean payoff larger than that of the resident ---regardless of their payoff variance--- can invade with a probability larger than the neural one. 

On the other hand, for $p_1\neq p_2$, i.e. non-symmetric or skewed distributions, the neutral line has a more complicated analytical form involving higher cumulants of the payoff distribution, such as the variance $\sigma^2$ and skewness (the latter encapsulated in the ratio $p_1/p_2$).  In particular, one finds the following scenarios (see Figure~\ref{fig:neutral_line}):

\begin{itemize}
\item If the distribution of payoffs is left-skewed ($p_1 < p_2$), so that the favorable environment is the more likely  one, the line of neutral selection tilts to the left in the $(\mu,\sigma)$ plane, as shown in Figure~\ref{fig:neutral_line}A. 
This indicates that increased payoff variance can lead to a higher invasion probability, even if the mean payoff is reduced, as illustrated by the red and green dots in the figure.
This illustrates that variance-prone strategies, specialized in exploiting the best environment, can be beneficial under these circumstances.

\item If the payoff distribution is right-skewed ($p_1 >p_2$), so that the unfavorable environment is the more likely one, the neutral line tilts to the right (Figure~\ref{fig:neutral_line}C). A higher invasion probability can thus be achieved by  reducing the payoff variance, even if the mean payoff is also reduced, see again the red and green dots in the figure. Therefore, it can be beneficial for a mutant to adopt a variance-averse strategy and become a generalist.

\end{itemize}

In summary, in the fast-switching limit, the level of asymmetry in the environmental distribution determines if there is room for risk-prone or risk-averse strategies. In simple terms: if the good environment is very likely, it may be beneficial to maximize the gains from it (increasing \(a_2\)) at the expense of reducing performance in the unlikely bad environment (lowering \(a_1\))  even if the mean payoff \(\mu\) decreases. Conversely, if the best environment is unlikely, it becomes advantageous to minimize risks by maximizing \(a_1\), even at the cost of reducing \(a_2\) and lowering the mean payoff. The first scenario leads to a variance-prone strategy, while the second results in a variance-averse approach.

\subsection{Slow-switching limit}\label{subsec: Ultra-slow switching}

In this limit, the environment changes so rarely that it can be considered fixed along any single evolutionary trajectory. This implies that the mean fixation probability of a mutant can be written as the average of the fixation probabilities across the possible fixed environmental conditions [Eq.~(\ref{eq:phi})],
\be\label{eq:phi_slow}
\phi_1=p_1\phi_{1,\omega=1}+p_2\phi_{1,\omega=2},
\ee
where:
\be\label{eq:phi_slow_2}
\phi_{1,\omega}=\frac{1-\gamma_\omega}{1-\gamma_\omega^N}, ~~~ \text{   with   }~~~ \gamma_\omega=\frac{1+\exp(-\Delta\pi_{\omega})}{1+\exp(\Delta\pi_\omega)}.
\ee
Recalling that the resident payoffs have been set to zero, one can read off $\Delta\pi_\omega$ from Eq.~(\ref{eq:inversion}) and find that $\Delta\pi_1=\mu-\sigma\sqrt{p_2/p_1}$  and $\Delta\pi_2=\mu+\sigma\sqrt{p_1/p_2}$. The line of neutral selection $\phi_1=1/N$ 
can be obtained analytically for large enough populations. Indeed, notice that for $\phi_1$ to be larger than zero  in the limit  $N\to\infty$, and taking into account that $p_1$ and $p_2$ are fixed and positive, it is sufficient if either $\phi_{1,\omega=1}$ or $\phi_{1,\omega=2}$ remain non-zero. 

In summary, the above result means that in the slow-switching limit and for sufficiently large populations, the mutant fixation probability is positive if the mutant has a payoff larger than the resident type in at least one of the environments. This favors variance-prone strategies maximizing the payoff under favorable conditions at any cost.

\subsection{Infinitely-large population size limit}

We now present some analytical findings for the case of infinitely large populations, i.e. for $N\rightarrow \infty$, independently of the environmental switching rate. In this case, the population follows  a deterministic replicator-like equation under any fixed environmental state $\omega$ (see e.g. \cite{traulsen_coevolutionary_2005,Sireci}):
\be
\dot x = T_\omega^+(x)-T_\omega^-(x),
\ee
with $T_\omega^\pm(x)$ obtained from the $T_{i,\omega}^\pm$ in Eq.~(\ref{eq:fermi}) by setting $x=i/N$ (and where a factor $N$ has been reabsorbed in the time scale, to make time continuous; i.e. one unit of time corresponds to one generation in the population). 
Including the dependence of the environmental state with time, leads to the replicator equation:
\be\label{eq:adjusted replicator}
\dot{x}(t)=x(t)\left[1-x(t)\right]\tanh\left[ \frac{\beta}{2} \Delta\pi_{\omega(t)} \right],
\ee
which is a piecewise-deterministic Markov process \cite{pdmp}, where the environment switches with rates $f_\pm=p_\pm/N$ and, between switches, the population follows the deterministic differential Eq. (\ref{eq:adjusted replicator}).

\begin{figure}
\centering
\includegraphics[width=0.5\textwidth]{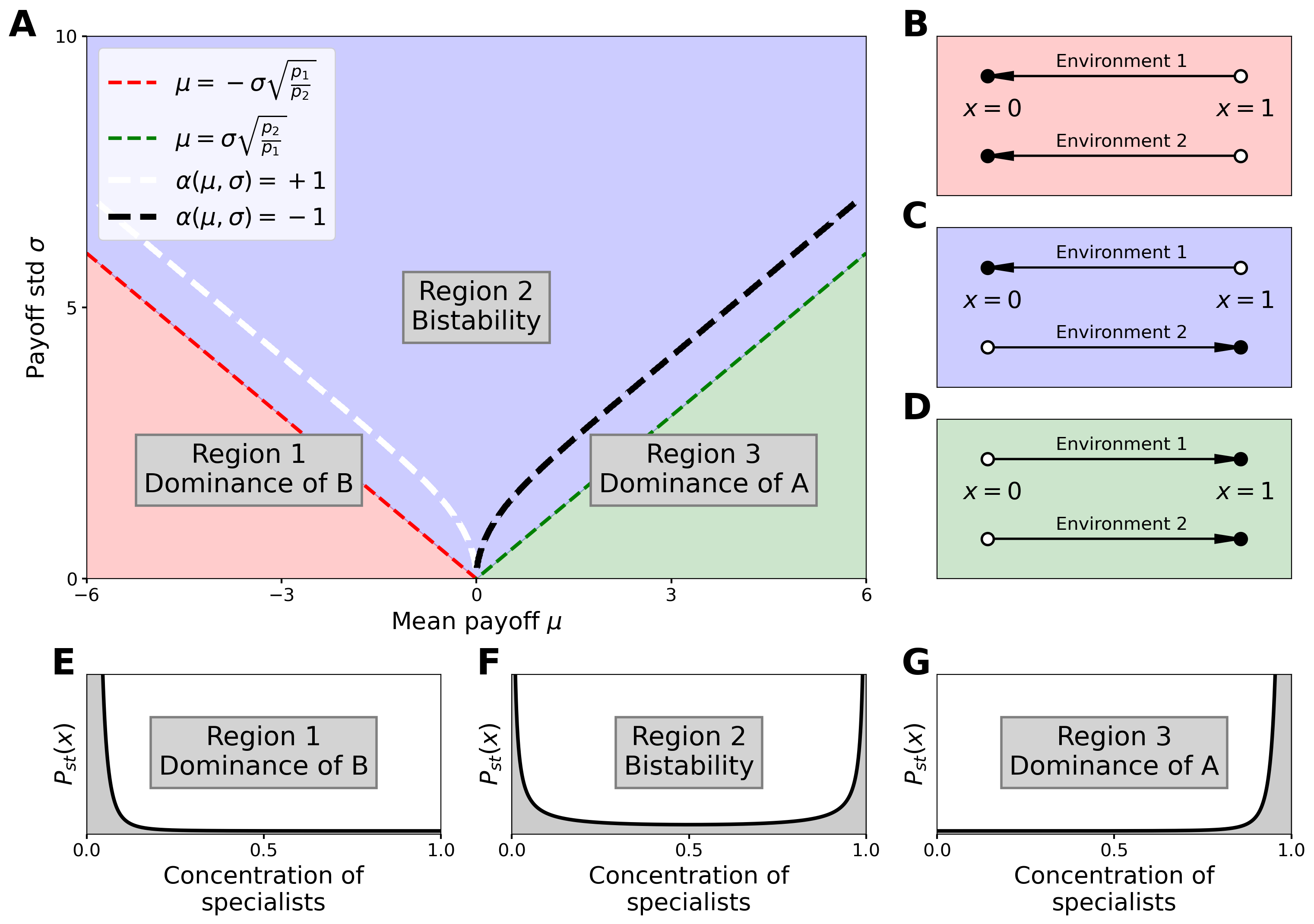}
        \caption{ {\bf Solution for infinite populations.} Panel A illustrates the behavior of the model as a function of the mean payoff $\mu$ and the standard deviation $\sigma$ of the mutant payoff for switching rates $f_\pm =1$ (and consequently, $p_1=p_2=1/2$). The red region shows $\mu<-\sigma$. In this case, the mutant is driven to extinction in both environments (see panel B). In the blue region ($-\sigma<\mu<\sigma$) one environment favors the mutant and the other favors the resident (panel C). In the green region ($\mu>\sigma$) the mutant is driven to fixation in both environments (panel D). Looking at the stationary distribution of Eq. (\ref{eq:adjusted replicator}), for an infinitely large population, there exist three distinct regions.
        {\bf Region 1} ($\alpha(\mu,\sigma)>1$) in which the stationary probability distribution diverges at $x=0$ (see panel E); the more likely state is the one dominated by strategy B. 
       {\bf Region 2} ($-1 < \alpha(\mu, \sigma) < 1$) is characterized by a bimodal stationary distribution (see panel F), where the population alternates between phases dominated by the mutant and the resident; the most likely outcome is the dominance of strategy B. {\bf Region 3} ($-1<\alpha(\mu,\sigma)$) in which strategy A dominates the dynamics, the stationary distribution is peaked at $x=1$ (see panel G). Crucially, the boundaries of the bistability region depend on the switching probabilities and not only on the payoffs in the different environments. }
\label{fig:flows_deterministic}
\end{figure}

One can distinguish the following cases:
\begin{enumerate}
\item If $\Delta\pi_\omega<0$ for both $\omega$'s, the payoff of the mutant is smaller than that of the resident type in both environments. This happens whenever $\mu<-\sigma\sqrt{p_1/p_2}$, as shown in Figure \ref{fig:flows_deterministic} (red shaded region in panel A, with the deterministic flows shown in panel B). In this case, the mutant is deterministically driven towards extinction.

\item If $\Delta\pi_\omega>0$ for both $\omega$, the mutant's payoff is larger than the resident's in both environmental states. This is the case when $\mu>\sigma\sqrt{p_2/p_1}$, as shown in Figure \ref{fig:flows_deterministic} 
(green shaded region in panel A, with the deterministic flows shown in panel D). In this case, the mutant is deterministically  driven towards fixation. 
    
\item The mutant is favored in one environment but disfavored in the other (i.e. there are opposite 
deterministic flows in the two environments as shown in panel C) defining a region of bistability (blue shaded area Figure~\ref{fig:flows_deterministic}A):
\be\label{eq:interesting region}
-\sigma\sqrt{\frac{p_1}{p_2}}\leq \mu\leq \sigma\sqrt{\frac{p_2}{p_1}}.
\ee 
In this case, one cannot ascertain \emph{a priori} the fate of the mutant as it may depend on other factors such as the speed of environmental change as we show next.
\end{enumerate}

To address the third scenario, we determine, as detailed in Appendix \ref{app:intermediate}, the stationary probability distribution describing the odds of finding the stochastic process defined by Eq. (\ref{eq:adjusted replicator}) in state $x$ \cite{hufton_intrinsic_2016}: 
\be\label{eq:stationary probability}
P_{\rm st}(x)={\cal N} (1-x)^{\alpha-1}  x^{-(1+\alpha)}
\ee
where ${\cal N}$ is normalization factor and $\alpha$ is a constant that depends on the mean payoff $\mu$, the payoff variance, $\sigma^2$, and the switching rates $f_\pm$,
\be\label{eq:alpha exponent}
\alpha=\frac{f_+}{\tanh\left(\frac{\beta}{2}\left(\mu-\sigma\sqrt{\frac{p_2}{p_1}}\right)\right)}+\frac{f_-}{\tanh\left(\frac{\beta}{2}\left(\mu+\sigma\sqrt{\frac{p_1}{p_2}}\right)\right)}.
\ee
The coefficient $\alpha$ determines the behavior of the stationary distribution near $x=0$ and near $x=1$. In particular:
\begin{enumerate}
 \item  For $\alpha>1$, $P_{\rm st}(0)\rightarrow \infty$ and $P_{\rm st}(1)=0$ and the population is dominated by the resident $B$ (Figure~\ref{fig:flows_deterministic}E).
 
    \item For $-1<\alpha<1$,  $P_{\rm st}(0)\rightarrow\infty$ and $P_{\rm st}(1)\rightarrow \infty$. Hence the system shows bistability between mutant-dominated and resident-dominated populations (Figure~\ref{fig:flows_deterministic}F).
 
    \item For $\alpha<-1$, $P_{\rm st}(0)=0$ and $P_{\rm st}(1)\rightarrow \infty$, so that the  population is invaded by the mutant $A$ (Figure~\ref{fig:flows_deterministic}G).
\end{enumerate}
Therefore, for fixed switching rates $f_\pm$, one can determine the lines in the $(\mu,\sigma)$ plane separating regions with a dominance of $A$, bistability, and dominance of $B$, respectively by solving $\alpha(\mu,\sigma)=\pm 1$. The numerical solutions of these two limiting curves, obtained for some specific rates $f_\pm$, are shown as black and white dashed lines in Figure \ref{fig:flows_deterministic}A (see also Appendix~\ref{app:intermediate some limits}).
 
In summary, the analytical insights gained from the limit of an infinitely large population size reveal that, in addition to the relative probabilities of the environments, the actual time scale of environmental dynamics plays a crucial role in determining the fate of mutants.

\section{Intermediate switching times and finite population sizes}

We now analyze environments with intermediate switching rates and finite population sizes. The idea is to see to what extent the results derived analytically in previous sections also apply to other regimes less accessible to analytical treatment. In particular, our goal is to analyze the influence of  environmental timescales and the shape of the environmental-state distribution on possible emerging strategies. For this, we will rely on direct computational analyses as well as on numerical solutions of the full set of coupled equations, Eq. (\ref{eq:S1}) and  Eq. (\ref{eq:S2}) to determine the fixation probability in various regimes by working with relatively small population sizes ($N=50$ and $N=100$).

\subsection{Influence of environmental-switching timescales}

\begin{figure*}[htbp]
\centering
    \includegraphics[width=1.0\textwidth]{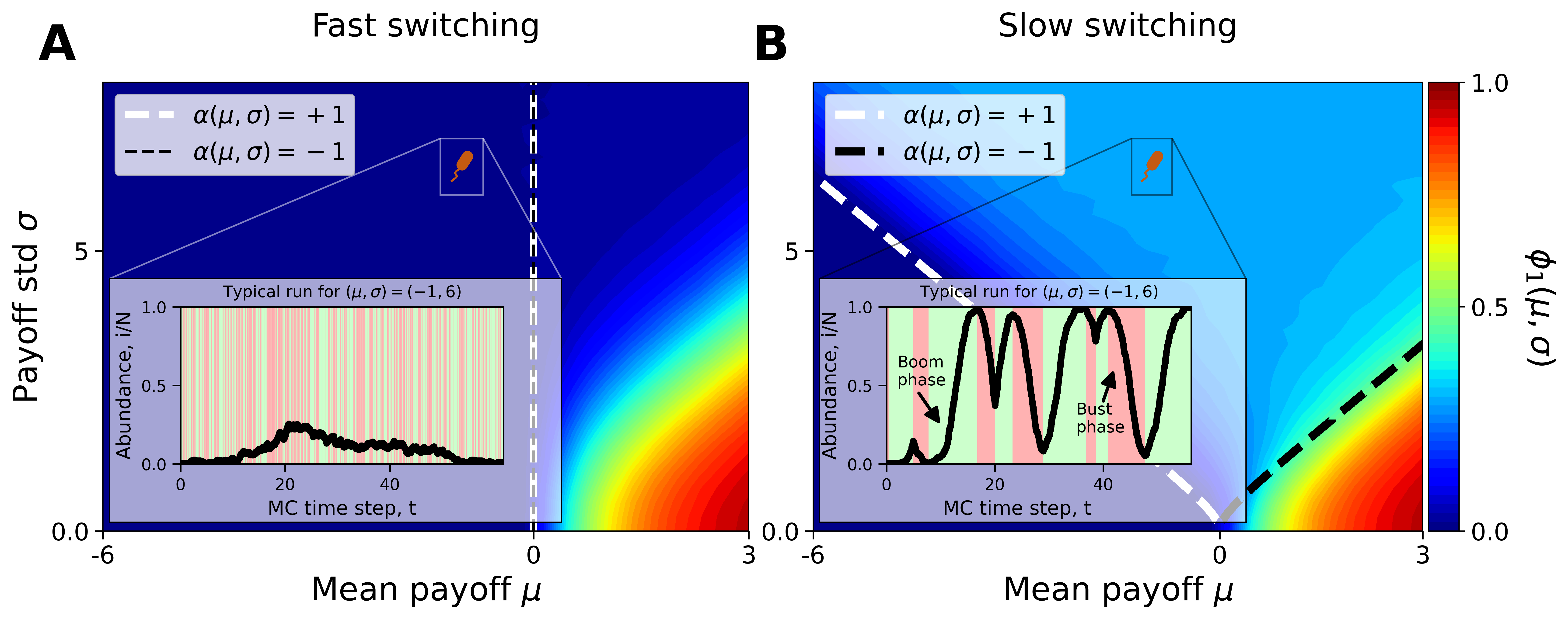}
   \caption{{\bf Influence of switching rates on fixation for symmetric environmental distributions.} Both panels depict the fixation probability $ \phi_1 $ as a function of the payoff mean $ \mu $ and standard deviation $ \sigma $ for a population of $ N = 100 $ individuals, with a selection strength $ \beta = 1 $, and $ p_+ = p_- $ (symmetric environment). Results are averaged over $ M = 10^5 $ independent runs. 
Panel A is for $ p_+ = p_- = 1 $ (fast switching), and panel B is for $ p_+ = p_- = 0.1/N $ representing slow switching.
Insets illustrate the fate of a mutant with a mean fitness $ \mu = -1 $ and standard deviation $ \sigma = 6 $. For slow switching (panel B) the mutant undergoes boom and bust cycles. Windows of opportunity emerge during favorable seasons, and the mutant has the potential to experience significant growth and reach fixation. If adverse conditions persist for too long, the mutant can go extinct. The white and black lines indicate the onset of bistability in infinite populations (see text for details).}
    \label{fig: Influence of the time scale}
  \end{figure*}

Figure~ \ref{fig: Influence of the time scale} shows the fixation probability, $ \phi_1(\mu,\sigma)$ (colored background) in a finite population, $N=100$, as a function of the mean $\mu$ and standard deviation $\sigma$ of the mutant payoff for two different choices of environmental-switching rates, but with $p_1=p_2$ in both cases. In particular, Panel A corresponds to the case where $ p_+ = p_- = 1 $, i.e., the environment switches state after each (attempted) birth-death event in the population. Panel B shows the case $ p_+ = p_- = 0.1/N $, representing a slow-switching scenario, where the environment changes, on average, once every ten generations. The white and black dashed lines ($ \alpha = 1 $ and 
$ \alpha = -1 $, respectively) mark the limit of the bistability region as determined analytically above for the limit of large population sizes. Observe that the region of bistability vanishes in the ultra-fast limit, as discussed in what follows, and broadens 
upon approaching the ultra-slow switching limit.

The insets show representative trajectories of the proportion of mutants in the population; the mutant in this example has been chosen to be inferior to the resident on average ($\mu=-1$) and to  have  a payoff standard deviation $\sigma = 6$, implying that it performs significantly better in the preferred environment than in the other (i.e., it is quite specialized).

In the fast-switching regime (Figure \ref{fig: Influence of the time scale}A) the lines $\alpha=\pm 1$ collapse into a single one determined by $\mu=0$, in agreement with the results for general $N$ in Section~\ref{sec:fast_switching} (see also Appendix \ref{app:intermediate some limits} for an analytical demonstration of this). Thus, a mutant with a mean payoff lower than the resident's is unlikely to invade the wild-type population. This is because the environment switches so rapidly that the mutant experiences an effective average environment, unfavorable for $\mu<0$ (and favorable only if it had a larger mean, $\mu>0$).

On the other hand, when the environment switches more slowly (Figure \ref{fig: Influence of the time scale}B) the analytical solution predicts a region of bistability (the area between the white and black dashed lines). In this region, a mutant with the same mean payoff $\mu=-1$ and the same variance $\sigma=6$ as in panel A is much more likely to be able to take over the resident population. In this case, the mutant can take advantage of the temporal correlations of environmental noise. Indeed, as the environment switches slowly, the mutant evolves through a series of ``boom and bust cycles". During these cycles, the mutant capitalizes on favorable environmental conditions to experience significant growth (boom), followed by a collapse (bust) once the environment reverts to unfavorable conditions. Eventually, if one of the boom phases persists long enough, the mutant can successfully fixate (as illustrated by the  realization shown in the inset). Conversely, the mutant can also become extinct during a sufficiently long adverse period.

Thus, in summary, slowly changing environments open more space for specialized mutants to invade the population, even if they perform on average worse than the resident type, by exploiting particularly favorable (boom) periods in which they thrive.

\subsection{Influence of the  payoff-distribution skewness}

\begin{figure*}[htbp]
\centering
    \includegraphics[width=1.0\textwidth]{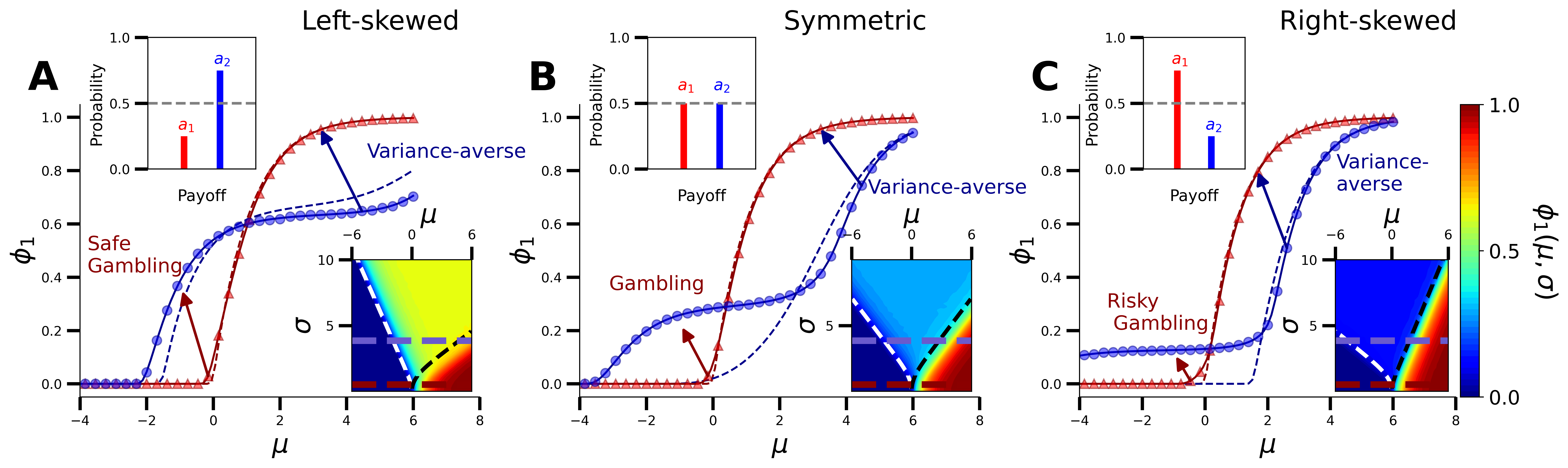}
   \caption{{\bf Effects of payoff asymmetry on mutant fixation.} Panels A, B, and C show mutant fixation probability for left-skewed, symmetric, and right-skewed payoff distributions ($p_1=0.25$, $p_1=0.5$, and $p_1=0.75$ respectively). The distributions are illustrated in the upper left insets. The main panels show the fixation probability as a function of mean mutant payoff $\mu$ for two different values of the standard deviation, $\sigma=1$ (red), and $\sigma=4$ (blue). This corresponds to the two cuts shown in the lower-right insets. Markers are from simulations of
    a population of size $N=50$ with an intensity of selection of $\beta=1$. Switching probabilities are $p_+=0.1/N, p_-=0.1/(3N)$ in panel A, $p_+=p_-=0.1/N$ in panel B, and $p_+=0.1/(3N), p_-=0.1/N$ in panel C. Data is averaged over $M=10^5$ realizations. The continuous lines show the theoretical predictions [Eqs.~(\ref{eq:S1}) and (\ref{eq:S2})], while the dashed lines in the main panels show the fast-switching approximation.}
  \label{fig:Asymmetry}
\end{figure*}

 As we have already seen in the limit of fast-switching environments (see Figure \ref{fig:neutral_line}), the relative frequencies of each of the two environmental states (determined by the ratio of $p_+$ and $p_-$ via Eq.(\ref{eq:p1p2})), control the possible emerging strategies. In what follows, we analyze the influence of the payoff asymmetry (or skewness) away from the fast-switching limit, i.e. for a slower, intermediate, switching $p_\pm\leq 0.1/N$ to see that novel possible strategies emerge.

In particular, Figure~\ref{fig:Asymmetry} illustrates outcomes for three distinct scenarios. In panel A, the mutant's most favorable environment is the most probable state (resulting in negative skewness). In panel B, both environments are equally likely (zero skewness), and in panel C the unfavorable environment is the more probable state  (positive skewness). The insets show the mutant fixation probability as a function of $ \mu $ and $ \sigma $, while the main plots represent two horizontal sections with small, $ \sigma = 1 $ (red),  and large $ \sigma = 4 $ (blue), payoff variance respectively. The continuous lines in the main plots show numerical solutions of Eqs.~(\ref{eq:S1}) and (\ref{eq:S2}), that can be explicitly calculated given the relatively small size of the population ($N=50$). 
Note that the analytical lines fit very well with simulated data (symbols). On the other hand, the dashed lines in the main plots of Figure~\ref{fig:Asymmetry} are predictions from the fast-switching approximation in Eq.~(\ref{eq: effective transition rates}). There are some discrepancies between the analytical curves derived for the fast-switching limit and the simulation results for slow-switching environments.

In particular, in all cases when $\mu<0$, i.e. if the mutant performs worse on average than the resident, there is space for variance-prone (or gambling) strategies (red arrows in Figure~~\ref{fig:Asymmetry}). The fixation probability can be enlarged by increasing the performance in the preferred environment, even if this comes at the price of a further decrease in its performance in the un-preferred environment; i.e. the mutant can extract a benefit by increasing its variance even at the expense of reducing its mean.

However, there is a slight difference between the two extreme cases reported in 
Figure~\ref{fig:Asymmetry}. For left-skewed payoff distributions ($p_2>p_1$), variance-prone or gambling can be labeled as ``safe", given that the environment in which the mutant specializes is the more likely one (Figure~\ref{fig:Asymmetry}A). On the other hand, for right-skewed distributions ($p_2<p_1$), these gambling strategies can be termed ``risky" as the preferred environment is the more unlikely one.

The second important difference between the cases with left-skewness and right-skewness is that --as can be observed by comparing with Figure \ref{fig:neutral_line}-- the risky-gambling regime only exists for sufficiently slow environmental changes as it disappears in the fast-switching regime. The reason for this is the same as that illustrated in Figure
\ref{fig: Influence of the time scale}: boom cycles need to be long enough to allow for the mutant specialized in the most infrequent environment to have the time to become fixated. As the rate of switching is decreased, the bistability region is enlarged and therefore there is more space for risky gambling.

In addition to these gambling  strategies, one can also identify variance-averse strategies (indicated by blue arrows in Figure~\ref{fig:Asymmetry}). Let us note that such strategies are advantageous for the mutant when it consistently outperforms the resident on average ($ \mu \gtrsim 0 $). In such scenarios, the optimal strategy for the invading mutant to dominate the resident generalist is to ``play it safe". This entails sacrificing some of its mean payoff in exchange for reduced environmental variability. A mutant employing this strategy thus becomes more of a generalist.

In summary, increasing payoff variance at the expense of diminished mean payoff is an effective strategy only under restricted conditions.  Negative asymmetry (left skew) facilitates this type of (safe) gambling as already discussed in the fast-switching limit. On the other hand, for symmetric or positively skewed environmental distributions gambling strategies can only be successful for sufficiently slow environments, and are inherently risky.

\section{Bet-hedging and fitness-based perspectives} \label{sec: real bet-hedging}
Thus far, we have discussed different strategies as a function of the distribution of payoffs for the mutant. However, to establish a direct link with ideas of traditional bet-hedging, we now shift the focus from payoffs to fitness.

As before, we focus on the case of two types of individuals, denoted $A$ and $B$ respectively. The conventional replicator equation is then given by
\begin{equation}\label{eq:replicator}
    \dot{x}(t)=x(t)(1-x(t))[f_A(t)-f_B(t)],
\end{equation}
where $f_A$ and $f_B$ represent the reproductive fitnesses of types $A$ and $B$, representing the relative increase in the number of individuals of either type over one generation,  at time $t$ \cite{hofbauer}. Comparing Eq.~(\ref{eq:replicator}) with Eq.~(\ref{eq:adjusted replicator}) for a fixed environment $\omega$, we have
\begin{equation}
    f_A-f_B=\tanh\left(\frac{\beta}{2} \Delta \pi\right),
\end{equation}
and one can make the identification
\BE
    f_A&=&g^+(\Delta \pi)=\displaystyle{\frac{1}{1+\exp(-\beta \Delta \pi)}}, \nonumber \\
    f_B&=&g^-(\Delta \pi)=\displaystyle{\frac{1}{1   +\exp(+\beta \Delta \pi)}},\label{eq:id}
\EE
establishing a non-linear relation between the fitnesses $f_A$ and $f_B$ and the payoff difference $ \Delta \pi $, as defined by Eq.~(\ref{eq:g_fermi}). 

We can thus re-evaluate our previous results as a function of the mean mutant fitness $ \mu_f $ and its standard deviation $\sigma_f $. 
In particular, writing $ f_{A,\omega}$ ($ \omega = 1,2$) for the fitness of the mutant strategy in environment $\omega$ we have
\be
\label{eq:inversion 2}
 f_{A,1} = \mu_f - \sigma_f \sqrt{\frac{p_2}{p_1}}, \;\;\; f_{A,2} = \mu_f + \sigma_f \sqrt{\frac{p_1}{p_2}}.
\ee
Note that the neutral case $\Delta \pi=0$ corresponds to $\mu_f=1/2$.

\begin{figure*}[htbp]
\centering
\includegraphics[width=1.0\textwidth]{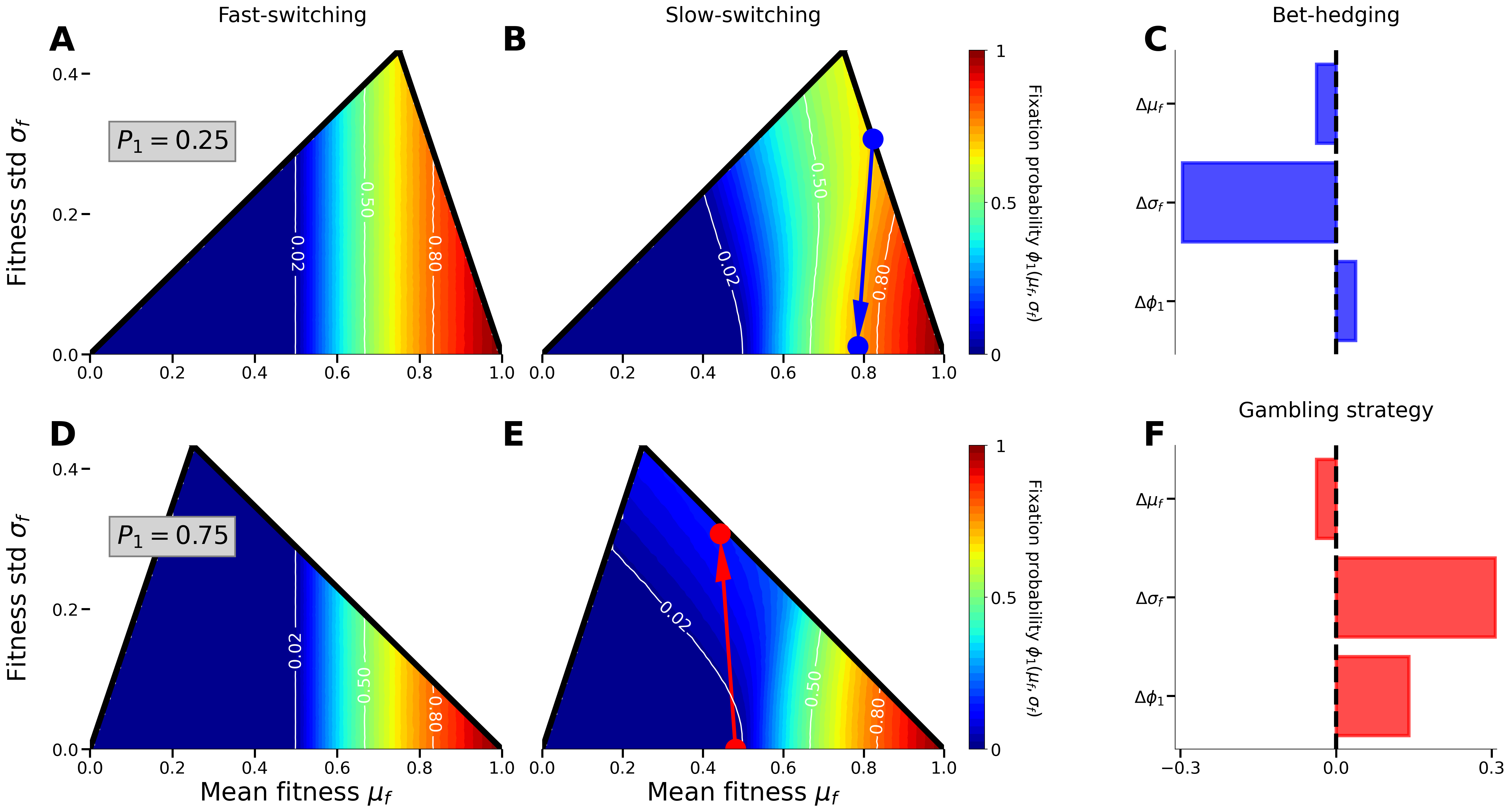}
   \caption{\textbf{Analyses in terms of fitness and \emph{bona fide} bet-hedging.}  
   The left column (panels \textbf{A} and \textbf{D}) represents fast switching, and the right column (panels \textbf{B} and \textbf{E}) is for slow switching. The first row (panels \textbf{A} and \textbf{B}) shows $p_1=0.25$, and the second row (panels \textbf{D} and \textbf{E}) is for $p_1=0.75$. 
   The first row, panel \textbf{A} and \textbf{B} illustrate the fixation probability as a function of mean fitness $ \mu_f $ and its standard deviation $ \sigma_f $ for the case where $ p_1 = 0.25 $ under fast and slow switching conditions, respectively.   Panels \textbf{D} and \textbf{E} present similar data, but with a probability $ p_1 = 0.75 $. Equal probability lines are depicted in white, with the smallest line ($ \phi_1 = 1/N = 0.02 $) denoting the line of neutral selection. 
   Fast switching here means $p_+=1$, $p_-=1/3$ for the case $p_1=0.25$ ---first row--- and $p_+=1/3$, $p_-=1$ for $p_1=0.75$ --second row-- while, on the other hand, slow switching means $p_+=0.1/N$, $p_-=0.1/(3N)$ for the case $p_1=0.25$ --first row-- and $p_+=0.1/(3N)$, $p_-=0.1/N$ in the case $p_1=0.75$ --second row--).
   The diagram is bounded by the values of $ \mu_f $ and $ \sigma_f $ such that $ f_A $ and $ f_B $ remain positive, delineating a triangular region.  Arrows in panels B and E denote, respectively, examples of bet-hedging and gambling strategies. Following the blue or red arrow results in an increase in fixation probability $ \Delta\phi_1 $ at the expense of a decrease in mean fitness $ \Delta\mu_f $ and a decrease (blue arrow) or increase (red arrow) of the standard deviation of fitness $ \Delta \sigma_f $, as illustrated in panels \textbf{C} and \textbf{F}. All simulations were conducted for a population size of $ N = 50 $ individuals, with a selection intensity of $ \beta = 1 $; averages were computed over $ M = 10^5 $ independent measurements.}
    \label{fig: real bet-hedging}
\end{figure*}

In Figure \ref{fig: real bet-hedging}, we present simulation results as a function of $\mu_f$ and $\sigma_f$ in various regimes. In particular,  we have $f_{A,\omega} \in[0,1]$ by construction (first relation in Eq.~(\ref{eq:id})) and, therefore, also $\mu_f\in[0,1]$. For a given value of $\mu_f$, Eq.~(\ref{eq:inversion 2}) then indicates that the standard deviation $\sigma_f$ is restricted to the interval $ \sigma_f \in [0, \min\{ \mu_f\sqrt{p_1/p_2}, (1-\mu_f)\sqrt{p_2/p_1} \}] $. These conditions define an allowed triangular region in $ (\mu_f, \sigma_f) $ space. 

Let us first recall that, in the fast-switching regime, the probability of fixation depends on payoffs solely via the ratio of the expected values of $ g^{\pm}(\Delta \pi)$  (see Eqs.~(\ref{eq: effective transition rates}) and (\ref{eq:phi_1_eff})). This coincides with the ratio of the average fitness of the two strategies. Consequently, the only relevant statistic in the fast-switching limit is the mean fitness $\mu_f$ of individuals of type $A$ across environments.  This means that, in the fast-switching limit, there is no room for strategies that are either averse or prone to fitness variance. The left column of Figure \ref{fig: real bet-hedging} (panels A and D) shows outcomes for the fast-switching limit and illustrates that the line of neutral selection is vertical at $ \mu_A = 1/2$ (resulting from $\Delta\pi=0$) and all other lines of constant fixation probability are vertical.

On the other hand, for slower switching environments, the line of neutral selection begins to curve leftward (Figure~\ref{fig: real bet-hedging}B and E). Consequently, there is a possibility  for gambling strategies, i.e., a strategy that increases the fitness variance while reducing the mean fitness across environments. In panel E, the unfavorable environment is the more likely state ($p_1=0.75$), so any gambling would be risky in the sense of the previous sections. An instance of such a strategy is illustrated by the red arrow. The increase in mutant fixation probability  is illustrated in panel F, along with the changes in mean mutant fitness across environments and in the standard deviation of fitness. 
 
Alternatively, when $ \mu_f > 1/2 $, some  lines of constant invasion probability bend towards the right in the $(\mu_f,\sigma_f)$ space (Figure \ref{fig: real bet-hedging}B and E), indicating the potential for bet-hedging strategies.

More specifically, the fixation probability can be increased by decreasing fitness variance, even at the expense of reducing mean fitness. This is illustrated by a blue arrow in panel B. We show the resulting increase in fixation probability and the associated reductions in mean fitness and standard deviation in panel C. This demonstrates that there is room for \emph{bona-fide} bet-hedging strategies, although the advantage is not very large in this example.

\section{Summary and conclusions}
We have studied the evolution of a finite population in randomly fluctuating external conditions. We focus on the lineage of a single mutant in a resident population; our main interest is in mutant strategies that reduce or increase payoff variation across environments to enhance the probability of success. Variance-averse strategies attempt to minimize variability across environments, thus avoiding worst-case scenarios. Variance-prone strategies, on the other hand, embrace risk and aim at maximizing the benefits of transient favorable conditions. We study if and when these strategies can increase the probability that the mutant reaches fixation. Depending on the circumstances, variance-averse or variance-prone strategies can enhance the mutant's success. This depends, in particular, on the distribution of payoffs across environments and on the time scales of environmental fluctuations.

We have developed an analytical theory when the environmental dynamics are so fast that the environment can effectively be averaged out, or so slow that no environmental changes occur before fixation, as well as in the limit of infinitely large populations.

In the (ultra) fast-switching limit, we find that variance-prone strategies can be successful for left-skewed payoff distributions (Figure~\ref{fig:neutral_line}), i.e., when the preferred environment is the more frequent one; in these cases, it becomes convenient to become specialized in the ``good'' environment even at the cost of reducing the performance in the ``bad'' one. Instead, when the payoff distribution is symmetric, the only way for a mutant with a mean payoff below the resident to enhance its chances of fixation above the neutral line is to increase its average payoff. Finally, variance-averse strategies tend to be most successful for right-skewed payoff distributions, where the preferred environment is rare.

On the other hand, in the limit of very slow switching, any evolutionary trajectory occurs in a fixed environment. Thus, the distinction between variance-prone and variance-averse strategies loses its significance. For large population sizes, analytical results show that the mutant can achieve fixation with a non-zero probability if its payoff exceeds that of the resident in at least one environment \cite{Traulsen_2008}. 
Thus, in this limit, it becomes beneficial to specialize in one of the environments.

For intermediate switching rates, partial analytical progress can be made in the limit of infinite populations. The system is then described by a piecewise deterministic Markov process, whose stationary distribution can be obtained. Doing this, we identify a region of bistability in the space spanned by the mean payoff $\mu$ and the payoff standard deviation $\sigma$. In this region, the population experiences alternating pulls towards mutant extinction and mutant fixation, and a series of ``boom and bust" cycles emerge. Along these cycles, the mutant may either become extinct or reach fixation in finite populations. The region of bistability increases when the environment changes more slowly, implying that there is more space for risk-prone strategies for more strongly time-correlated environments.

A detailed analysis shows that gambling is possible for intermediate environmental switching probabilities, provided the external circumstances are typically adverse, i.e., situations where the average mutant payoff is smaller than the resident's.
In this case, left-skewed payoff distributions promote safe gambling strategies (Figure ~\ref{fig:Asymmetry}A) as the preferred environment is the most frequent one. On the other hand, if the distribution is right-skewed, i.e. the preferred environment is less frequent, the mutant needs to gamble on the rare occurrence of the preferred environment, aiming at fixation during one of the boom phases. If external conditions are beneficial for the mutant (with average mutant payoffs exceeding resident payoffs), the optimal response is to "play it safe" by adopting a variance-averse strategy.

Conventional bet-hedging refers to strategies minimizing the variance of reproductive fitness rather than that of the payoff. It is therefore natural to ask if there is scope for fitness variance-averse (bet-hedging) and fitness variance-prone (gambling) strategies within our framework.

A superficial inspection of the fast-switching limit suggests that this might not be the case. Eq.~(\ref{eq:gamma}) in Appendix C indicates that the mutant's fixation probability under fast-switching environments is equal to that under neutral selection if and only if the mean mutant fitness is equal to that of the resident. However, for sufficiently slow switching and sufficiently low average mutant payoff, we find that there is indeed scope for strategies that increase fitness variation while at the same time reducing mean fitness (red arrow in Figure~\ref{fig: real bet-hedging}E). At the same time, \emph{bona-fide} bet-hedging strategies can be successful for higher average mutant payoffs.

The overall rationale for variance-averse behavior is simple: ``play it safe'', or ``don't put all your eggs in one basket''. However, the simplicity of this idea is deceiving, and it is well-recognized that there are many layers of complexity to the problem of determining the most appropriate response to uncertainty. For example, one may ask if  variance reduction occurs at the level of individuals or populations. This leads to the distinction between conservative and diversifying bet-hedging. It is important to bear in mind, though, that this distinction is less clear cut than often assumed \cite{Kokko}, and it may be more appropriate to think of a continuum of strategies.
Furthermore, it is perhaps not always clear what objective an organism tries to achieve. For example, is this long-term growth or, possibly, better short-term performance? This may lead one to ask what object’s variance needs to be reduced, and under what conditions risky opportunistic strategies might be more successful. Further complications involve the role of demographic noise; for instance, recent work \cite{weissman} suggests that stochasticity can reverse the direction of selection for a bet hedger. 

Our results contribute to the ongoing process of disentangling the intricacies of finding the most appropriate response to adversity and uncertainty. It may sometimes be easier and more relevant for an organism or gambler to consider short-term payoff rather than long-term reproductive fitness or wealth. We have found that this may determine whether a variance-prone or variance-averse strategy is preferable. As we have shown, the time scales of environmental change and the distribution of payoffs across environments also affect the answer to some of the questions above. Most markedly, we find that there can be circumstances in which risky gambling strategies are advised. This occurs when conditions are generally unfavorable, and positive opportunities are scarce.

We believe our work can help clarify the relation of variance-averse strategies, bet-hedging, and gambling, both in the face of uncertain payoff or fitness. Our findings suggest that opportunistic strategies may constitute a plausible choice for biological populations, complementing risk-averse strategies such as bet-hedging. Both types of strategies might serve to enhance biodiversity \cite{doebeli_boom-bust_2021}. More generally, we hope that models  such as ours, combining demographic noise and external fluctuations will prove useful for future work on the study of strategies to cope with uncertainty.

\begin{acknowledgments}
MAM and RC acknowledge the Spanish Ministry and Agencia Estatal de investigaci\'on (AEI) for financial support through Project of I+D+i Ref  PID2023-149174NB-I00 and PID2020-113681GB-I00 funded by MICIN/AEI/10.13039/501100011033. TG acknowledges partial financial support from the Agencia Estatal de Investigaci\'on and Fondo Europeo de Desarrollo Regional (FEDER, UE) under project APASOS (PID2021-122256NB-C21, PID2021-122256NB-C22), and the Mar\'ia de Maeztu program for Units of Excellence, CEX2021-001164-M funded by MICIU/AEI/10.13039/501100011033. We are also grateful to Lorenzo Fant and Jorge Hidalgo for valuable discussions. 
\end{acknowledgments}

\newpage

\onecolumngrid

\appendix

\section{Arithmetic versus geometric means} \label{app: Arithmetic vs geometric}
In the simplest models of bet-hedging, one considers a process in which the fitness of a given phenotype varies across generations $t=1,2,\dots$. To illustrate matters and following \cite{Lewontin1969}, we focus on the most basic multiplicative process, $N_{t+1}=r_t N_t$, where $N_t$ is the size of a population in generation $t$.  If the values of $r_t$ at different times are independent and identically distributed, then the mean population $\langle N_t\rangle$ at time $t$ is given by $\langle N_t \rangle=N_0 \times \langle r\rangle^t$. Thus, in this case, the expected population growth is governed by the arithmetic mean $\langle r \rangle$. 

\begin{figure*}[htbp!]
\centering
\includegraphics[width=0.8\textwidth]{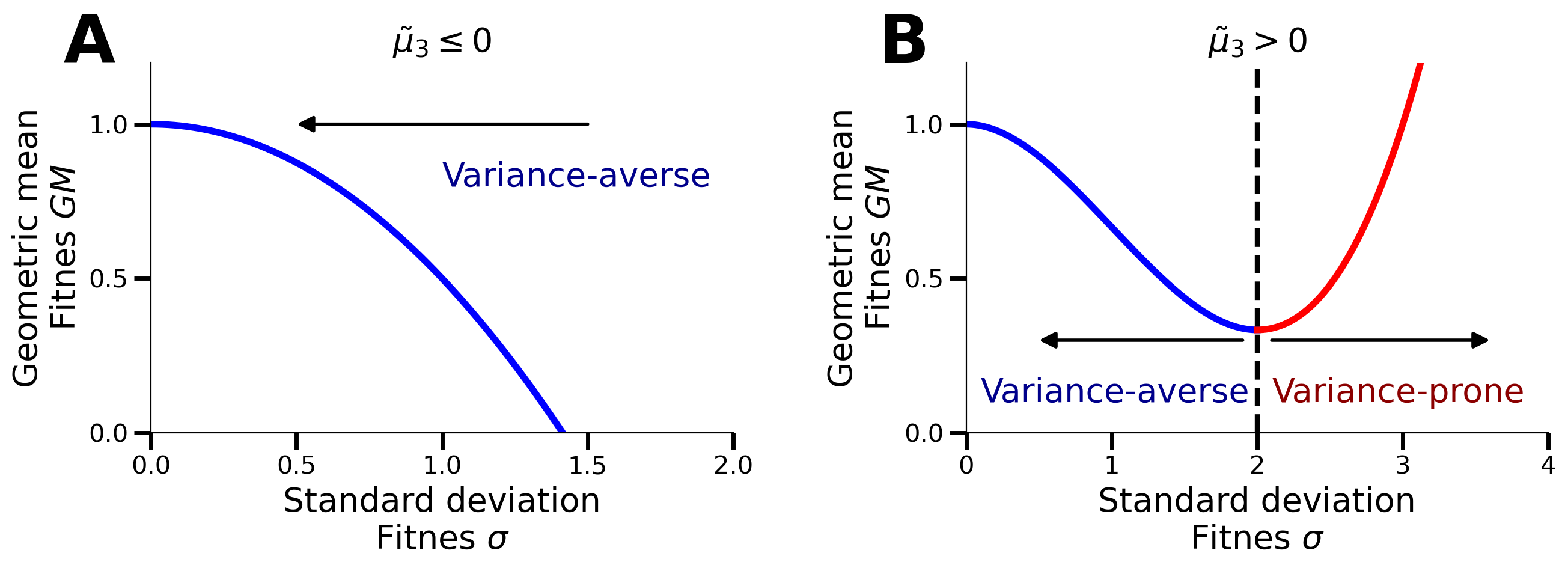}
   \caption{{\bf Geometric mean fitness as a function of the standard deviation of the fitness $\sigma$, given a mean value $\mu=1$}. Panel A represents the geometric mean for a non-positive skewness, while panel B shows the case where the skewness is positive. Panel A suggests that the strategy that maximizes the geometric mean fitness is one where the variance is reduced (variance-averse strategies). Panel B suggests that, in the case of positive skewness, the strategy that maximizes the geometric mean might be a variance-prone one. In particular, the dashed line shows a $\sigma$ value such that, increasing the variance is beneficial; for smaller values, a fitness gain is derived by slightly reducing the variance (variance-averse strategy).}
    \label{fig: Geometric mean}
\end{figure*}

Nonetheless, extinctions, for example, are controlled by the geometric mean of the random variable $r$, defined as the exponential of the average logarithmic growth rate:
\be 
GM=\exp(\langle \ln\,r\rangle).
\ee
Following \cite{Kokko}, one can expand this expression in terms of $\sigma$ (the standard deviation of the random variable $r$) by assuming that $r\approx \mu+\delta$, where $\mu$ is the arithmetic mean of $r$ and where the random variable $\delta=r-\mu$ is a relatively small deviation. Thus, a Taylor expansion to third order leads to the following expression:
\be \label{eq: GM approx}
GM\approx \mu-\frac{1}{2}\frac{\sigma^2}{\mu}+\frac{1}{3}\frac{\sigma^3\tilde{\mu}_3}{\mu^2}+o(\delta^4),
\ee
where $\tilde{\mu}_3=\langle\delta^3\rangle/\sigma^3$ is the skewness of the distribution of $r$. Interestingly, when the skewness is $0$ (or the variance of the distribution is sufficiently small so that higher-order terms can be neglected), the geometric mean is controlled by the mean $\mu$ and the variance $\sigma^2$. Taking into account only the first two terms in Eq.~(\ref{eq: GM approx}) it is always convenient to reduce the fitness variance to maximize the geometric mean (see Figure \ref{fig: Geometric mean}A). This is the basis of bet-hedging theory.

The situation changes however when one takes into account the third-order term. Indeed, Eq.~(\ref{eq: GM approx}) becomes a third-degree polynomial whose shape depends on the skewness. If the skewness is negative, reducing the variance is always beneficial, just as in the symmetric case (Figure \ref{fig: Geometric mean}A). If, on the other hand, the skewness is positive, the third-degree polynomial has the shape depicted in Figure \ref{fig: Geometric mean}B, indicating that increasing the variance might turn out to be beneficial in some scenarios.

\section{Fixation probabilities in the fast-switching limit}\label{app:fixation_probs}
As in the main text, we consider a fixed-size population with $N$ individuals, each of which can be in one of two possible states: $A$ or $B$. The probability of finding the population in a state with $i$ individuals of type $A$ and $N-i$ individuals of type B at time $t$ is denoted as $p_i(t)$. We focus on the ultra-fast-switching limit in discrete time, in which the environment $\omega$ is drawn at random from the stationary distribution of the environmental process before each birth-death event. For a given state $\omega$ of the environment, one has
\be\label{eq:master_app}
p_{i|\omega}(t+1)=(1-T_{i,\omega}^+-T_{i,\omega}^-)p_i(t)+p_{i-1}(t)T_{i,\omega}^+p_{i-1}(t)+p_{i+1}(t)T_{i,\omega}^-p_{i+1}(t),
\ee
where $p_{i|\omega}(t+1)$ is the probability to find the system in state $i$ at time $t+1$ {\em given} that an environmental state $\omega$ was drawn in the time step from $t$ to $t+1$.

The probability of finding the system in state $i$ at time $t+1$ (without conditioning on the environmental state) is then
\be
p_i(t+1)=\sum_{\omega} \, \rho(\omega) p_{i|\omega}(t).
\ee
Averaging both sides of Eq.~(\ref{eq:master_app}) over $\omega$, we therefore have
\be
p_{i}(t+1)=(1-T_{i,{\rm eff}}^+-T_{i,{\rm eff}}^-)p_i(t)+p_{i-1}(t)T_{i,{\rm eff}}^+p_{i-1}(t)+p_{i+1}(t)T_{i,{\rm eff}}^-p_{i+1}(t),
\ee
The time evolution of the probabilities $p_i(t)$ is therefore described by a master equation with the effective rates $T^\pm_{i,{\rm eff}}=\sum_\omega \rho(\omega)T_{i,\omega}^\pm$. These are the transition rates averaged over the possible environmental states.

\section{Neutral selection in the fast-switching limit}\label{app:neutral_line}
The relevant quantity in the (ultra) fast switching limit is 
\be
\label{eq:gamma}
\gamma_{\rm eff}=\frac{\int da \, \rho(a) g^-(a)}{\int da \,\rho(a) g^+(a)},
\ee
where $\rho(a)$ is the distribution of mutant payoffs across environments. More precisely, the neutral line is determined by the condition $\gamma=1$, that is
\be
\int da\, \rho(a) g^-(a)=\int da \, \rho(a) g^+(a).
\ee
Given that $g^+(a)+g^-(a)$ for all $a$, this means 
$\int da\, \rho(a) g^-(a)=\int da \, \rho(a) g^+(a)=1/2$.
In the simple case of a dichotomous distribution $\rho(a)=p_1\delta(a-a_1)+p_2\delta(a-a_2)$, with $p_1+p_2=1$ we find the condition
\begin{equation}
\label{eq:I_minus_app}
\frac{p_1}{1+\exp\left(\displaystyle{-\beta\left[\mu-\sigma\sqrt{\frac{p_2}{p_1}}\right]}\right)}+\frac{p_2}{1+\exp\left(\displaystyle{-\beta\left[\mu+\sigma\sqrt{\frac{p_1}{p_2}}\right]}\right)}=\frac{1}{2}.
\end{equation}
We now write $\mu$ along the neutral line in the following form [with $f(\sigma,p_1)$ to be determined],
\begin{equation}\label{eq:mu_app}
\mu=-\sigma\sqrt{\frac{p_1}{1-p_1}}-\frac{1}{\beta}\ln\,f(\sigma,p_1).
\end{equation}
Using this in Eq.~(\ref{eq:I_minus_app}) one then finds
\BE
&&-2\Gamma^3(\sigma,p_1)\exp\left(-2\beta\sigma\sqrt{\frac{p_2}{p_1}}\right)f^2(\sigma,p_1)+f(\sigma,p_1)(p_2-2)\Gamma(\sigma,p_1)+f(\sigma,p_1)(p_1-2)\nonumber \\
 &&+f(\sigma,p_1)(p_1-2)\Gamma^2(\sigma,p_1)\exp\left(-2\beta\sigma\sqrt{\frac{p_2}{p_1}}\right)f^2(\sigma)+(p_1+p_2-2)=0,
\EE
with \begin{equation}\label{eq:gamma_app}
\Gamma(\sigma,p_1)=\exp\left[\beta\sigma\left(\sqrt{\frac{p_1}{1-p_1}}+\sqrt{\frac{1-p_1}{p_1}}\right)\right].
\end{equation}
For given values of $p_1, p_2$ and $\beta$ this is a quadratic equation in $f(\sigma,p_1)$. After removing the negative solution [which would not be meaningful inside the logarithm in Eq.~(\ref{eq:mu_app})] we are left with:
\begin{equation}\label{eq:def_f}
f(\sigma,p_1)=\left(p_1-\frac{1}{2}\right)\left(\frac{1}{\Gamma(\sigma,p_1)}-1\right)+\frac{1}{2\Gamma(\sigma,p_1)}\sqrt{4(\Gamma(\sigma,p_1)-1)^2p_1(p_1-1)+(\Gamma(\sigma,p_1)+1)^2},
\end{equation}

We have therefore derived the set of Eqs.~(\ref{eq:neutral_line}), (\ref{eq:f}) and (\ref{eq:gamma_main}).

\section{Neutral selection in the intermediate switching regime}\label{app:intermediate}
We focus on the piecewise deterministic process in Eq.~(\ref{eq:adjusted replicator}), where the environment switches states with rates $f_\pm$. Using the results from \cite{hufton_intrinsic_2016}, we can compute the stationary probability distribution as
\begin{equation}
    P_{st}(x)=\mathcal{N}_1\left( p_1 \frac{h(x)}{-u_1(x)}+p_2\frac{h(x)}{u_2(x)}\right), 
\end{equation}
where $\mathcal{N}_1$ is a normalization constant, $h(x)$ is the function
\begin{equation}
    h(x)=\exp\left[ -\int^x \left( \frac{f_+}{u_1(s)}+\frac{f_-(s)}{u_2(s)} \right)ds \right],
\end{equation}
 and the $u_\omega(x)$ are the deterministic flows in the two environments [the right-hand side of Eq.~(\ref{eq:adjusted replicator})], $u_\omega(x)=x(1-x)\tanh\left(\beta\Delta\pi_\omega/2\right)$. After some algebra, one arrives at the following expression for the stationary probability distribution:
\be
P_{\rm st}(x)={\cal N}_2 \frac{(1-x)^{\alpha-1}}{x^{\alpha
+1}},
\ee
where ${\cal N}_2$ ensures normalization, and where the parameter $\alpha$ is given by:
\be \label{eq: app alpha}
\alpha=\frac{f_+}{\tanh\left(\frac{\beta}{2}\left(\mu-\sigma\sqrt{\frac{p_2}{p_1}}\right)\right)}+\frac{f_-}{\tanh\left(\frac{\beta}{2}\left(\mu+\sigma\sqrt{\frac{p_1}{p_2}}\right)\right)}
\ee

\section{Ultra-fast and slow switching limits of the piecewise deterministic process}\label{app:intermediate some limits}
We now study the limits of (ultra) fast switching, and slow switching of the piecewise deterministic process in Eq.~(\ref{eq:adjusted replicator}), with environmental switching rates $f_\pm$. The boundaries of the bistability region are given by the conditions $\alpha(\mu,\sigma)=\pm 1$.

\subsection{Limit of (ultra) fast switching}
First, setting $\alpha=\pm 1$ in Eq.  (\ref{eq:alpha exponent}) and subsequently dividing by $f_+$ on both sides gives:
\be
\frac{1}{\tanh\left(\frac{\beta}{2}\left(\mu-\sigma\sqrt{\frac{p_2}{p_1}}\right)\right)}+
\frac{f_-/f_+}{\tanh\left(\frac{\beta}{2}\left(\mu+\sigma\sqrt{\frac{p_1}{p_2}}\right)\right)}=\pm \frac{1}{f_+}.
\ee
Using Eq. (\ref{eq:p1p2}), we have $f_-/f_+=p_1/p_2$. Hence, we arrive at 
\be\label{eq: G2}
\frac{p_2}{\tanh\left(\frac{\beta}{2}\left(\mu-\sigma\sqrt{\frac{p_2}{p_1}}\right)\right)}+
\frac{p_1}{\tanh\left(\frac{\beta}{2}\left(\mu+\sigma\sqrt{\frac{p_1}{p_2}}\right)\right)}=\pm \frac{p_2}{f_+}.
\ee
Taking the limit $f_+\rightarrow \infty$ (at fixed $f_-/f_+=p_1/p_2$) the right-hand-side tends to zero, indicating that the two equations (for $\alpha=+1$ and one for $\alpha=-1$) reduce to a single relation
\be\label{eq:help}
-p_1\tanh\left(\frac{\beta}{2}\left(\mu-\sigma\sqrt{\frac{p_2}{p_1}}\right)\right)=p_2\tanh\left(\frac{\beta}{2}\left(\mu+\sigma\sqrt{\frac{p_1}{p_2}}\right)\right)
\ee
Next, we use 
\be
\tanh\left(\frac{\beta}{2} x \right)=g^+(x)-g^-(x),
\ee
with $g^\pm(x)$ defined as in the main text [see Eq. (\ref{eq:g_fermi})], the fact that $g^-=1-g^+$ and $p_1+p_2=1$. Eq.~(\ref{eq:help}) can then be written as
\be\label{eq:help2}
\frac{p_1}{1+\exp\left(\displaystyle{-\beta\left[\mu-\sigma\sqrt{\frac{p_2}{p_1}}\right]}\right)}+\frac{p_2}{1+\exp\left(\displaystyle{-\beta\left[\mu+\sigma\sqrt{\frac{p_1}{p_2}}\right]}\right)}=\frac{1}{2},
\ee
which is precisely the equation for the line of neutral selection in the fast-switching limit (see Eq.
 (\ref{eq:I_minus_app})). 

In the case of symmetric ultra-fast switching ($f_+=f_-\to\infty$) we have $p_1=p_2=1/2$. After some minor algebra, Eq.~(\ref{eq:help2}) then reduces to $\mu=0$.
\subsection{Slow-switching limit}
Returning to Eq. (\ref{eq: G2}) and taking the slow-switching limit $f_+\rightarrow 0$ (again at fixed $f_-/f_+=p_1/p_2$) we observe that the right-hand-side of Eq. (\ref{eq: G2}) diverges to $\pm\infty$. This means that the left-hand side must also diverge. This in turn implies that the argument of at least one of the hyperbolic tangent terms must tend to zero $0$, which leads to the conditions $\mu=-\sigma\sqrt{p_1/p_2}$ and $\mu=\sigma\sqrt{p_2/p_1}$. 

Thus, the entire region in $(\mu,\sigma)$ space in which the mutant is favored in one environment but disfavored in the other [defined by Eq. 
 (\ref{eq:interesting region}) of the main text] is now characterized by bistability. 

In particular, the boundary between the region in which the mutant is driven to extinction in both environments (red region in Figure ~\ref{fig:flows_deterministic}) and the one with bi-stability (blue) is given by $\mu=-\sigma\sqrt{p_1/p_2}$ for slow switching. In the former region, the invasion probability is zero, in the latter it is non-zero. The line $\mu=-\sigma\sqrt{p_1/p_2}$ is thus also the line of neutral selection in the slow-switching limit.

\end{document}